\documentclass{pspum-l}

\usepackage{color}
\definecolor{niceblue}{rgb}{0,0,0.6}
\usepackage[a4paper,colorlinks,citecolor=niceblue,linkcolor=niceblue,urlcolor=niceblue,pdfpagemode=None]{hyperref}
\usepackage{latexsym}
\usepackage{color}
\usepackage[all,cmtip]{xy}

\usepackage{stmaryrd}
\usepackage{amsmath, amscd, amssymb, mathrsfs, accents, amsfonts}
\usepackage{url}
\usepackage[all]{xy}
\usepackage{longtable}
\usepackage{dsfont}
\usepackage{tikz}
\usetikzlibrary{decorations.pathmorphing}
\usetikzlibrary{calc}
\usetikzlibrary{decorations.markings}

\def\nicecolourscheme{\shadedraw[top color=blue!22, bottom color=blue!22, draw=white]}

\def\nicepalecolourscheme{\shadedraw[top color=blue!22, bottom color=blue!22, draw=white]}

\definecolor{Myblue}{rgb}{0,0,0.6}

\makeatletter
\newcommand{\raisemath}[1]{\mathpalette{\raisem@th{#1}}}
\newcommand{\raisem@th}[3]{\raisebox{#1}{$#2#3$}}
\makeatother

\newcommand{\E}{\text{e}}
\newcommand{\I}{\text{i}}
\newcommand{\C}{\mathds{C}}

\newcommand{\be}{\begin{equation}}
\newcommand{\ee}{\end{equation}}
\newcommand{\bes}{\begin{equation*}}
\newcommand{\ees}{\end{equation*}}

\newcommand{\Hom}{\operatorname{Hom}}
\newcommand{\End}{\operatorname{End}}

\newcommand{\im}{\operatorname{im}}

\newcommand{\Hcc}{\mathcal{H}_{\text{(c,c)}}^A}
\newcommand{\Hrr}{\mathcal{H}_{\text{RR}}^A}

\newcommand{\hmf}{\operatorname{hmf}}

\newcommand{\ev}{\operatorname{ev}}
\newcommand{\tev}{\widetilde{\operatorname{ev}}}
\newcommand{\coev}{\operatorname{coev}}
\newcommand{\tcoev}{\widetilde{\operatorname{coev}}}
\def\lra{\longrightarrow}
\def\lmt{\longmapsto}
\DeclareMathOperator{\tr}{tr}

\numberwithin{equation}{section}
\newtheorem{punchline}{Punchline}
\newtheorem{example}{Example}
\newtheorem*{acknowledgements}{Acknowledgements}

\def\coev{\operatorname{coev}}

\tikzset{
    string/.style={draw=#1, postaction={decorate}, decoration={markings,mark=at position .51 with {\arrow[draw=#1]{>}}}},
    costring/.style={draw=#1, postaction={decorate}, decoration={markings,mark=at position .51 with {\arrow[draw=#1]{<}}}},
    ostring/.style={draw=#1, postaction={decorate}, decoration={markings,mark=at position .47 with {\arrow[draw=#1]{>}}}},
    ustring/.style={draw=#1, postaction={decorate}, decoration={markings,mark=at position .56 with {\arrow[draw=#1]{>}}}},
    oostring/.style={draw=#1, postaction={decorate}, decoration={markings,mark=at position .43 with {\arrow[draw=#1]{>}}}},
    uustring/.style={draw=#1, postaction={decorate}, decoration={markings,mark=at position .59 with {\arrow[draw=#1]{>}}}},
    directed/.style={string=blue!50!black}, 
    odirected/.style={ostring=blue!50!black}, 
    udirected/.style={ustring=blue!50!black}, 
    oodirected/.style={oostring=blue!50!black}, 
    uudirected/.style={uustring=blue!50!black},     
    redirected/.style={costring= blue!50!black},
    redirectedgreen/.style={costring= green!50!black},
    directedgreen/.style={string= green!50!black},
    redirectedlightgreen/.style={costring= green!65!black},
    directedlightgreen/.style={string= green!65!black},
}

\tikzset{-dot-/.style={decoration={
  markings,
  mark=at position 0.5 with {\fill circle (1.875pt);}},postaction={decorate}}}

\tikzset{
	Fdot/.style={circle, draw, fill, inner sep=0pt}, 
	Odot/.style={circle, draw, inner sep=0.1pt, minimum size=0.1cm}
	}

\def\nicecolourscheme{\shadedraw[top color=blue!22, bottom color=blue!22, draw=blue!22]}
\def\nicepalecolourscheme{\shadedraw[top color=blue!12, bottom color=blue!12, draw=white]}

\allowdisplaybreaks

\begin{document}

\title{\mbox{A quick guide to defect orbifolds}}

\author{Ilka Brunner}
\address{Arnold Sommerfeld Center for Theoretical Physics, LMU M\"unchen}
\email{ilka.brunner@physik.uni-muenchen.de}

\author{Nils Carqueville}
\address{Simons Center for Geometry and Physics}
\email{nils.carqueville@scgp.stonybrook.edu}

\author{Daniel Plencner}
\address{Arnold Sommerfeld Center for Theoretical Physics, LMU M\"unchen}
\email{daniel.plencner@physik.uni-muenchen.de}


\begin{abstract}
We provide a lightning review of the construction of (generalised) orbifolds \cite{ffrs0909.5013, genorb} of two-dimensional quantum field theories in terms of topological defects, along the lines of \cite{BCP}. This universal perspective has many applications, some of which we sketch in the examples of $2d$ Yang-Mills theory, Landau-Ginzburg models, and rational CFT. 
\end{abstract}

\maketitle

\section{Orbifolds via defects}
\label{sec:orbifoldsviadefects}

Orbifolds of two-dimensional quantum field theories arise from a standard construction \cite{dhvw1985, dvvv1989}. Starting from a theory with bulk space of states $H_e$ and a finite symmetry group~$G$ acting on it, one first considers the twisted sector spaces $H_g$ for each $g\in G$. In string theoretic terms their elements are strings that close only up to the action of~$g$. In general they are vertex operators~$\phi$ with the property $\phi(\E^{2\pi\I} \sigma) = g \cdot \phi(\sigma)$; in particular $H_e$ with $e\in G$ the unit is the original untwisted sector. To obtain the true state space $\mathcal H$ of the orbifold theory from $H = \bigoplus_{g\in G} H_g$, in a second step one has to average over all group actions:
$$
\mathcal H = P H
\, , \quad
P = \sum_{g\in G} P_g 
\, , \quad 
P_g = \frac{1}{|G|} \sum_{h\in G} h \big|_{H_g} \, . 
$$

A similar construction holds in the boundary sector: every boundary condition of the orbifold theory comes with a representation of~$G$, 
and boundary operators must be equivariant with respect to the induced representation. 

\medskip

Our first objective is to present a natural reformulation of the standard orbifold construction in terms of topological defects. 
Recall that defects are generalisations of boundary conditions in the sense that the latter are special cases of the former: boundary conditions are defects between the trivial theory and a given non-trivial theory. Also, by the `folding trick' defects in any theory (including orbifold theories) are treated in complete analogy to boundary conditions. For relevant background on topological defects we refer e.\,g.~to~\cite{Petkova:2000ip, Frohlich:2006ch, br0707.0922, dkr1107.0495}. 

Every unorbifolded theory has a distinguished defect, namely the invisible defect~$I$ which does not impose any conditions on bulk fields and acts as the identity under fusion. Furthermore for every $g\in G$ there is a defect ${}_g I$ which by construction implements the action of~$g$, i.\,e.~pulling ${}_g I$ across a field insertion~$\phi$ one obtains its image under $g$: 
$$
\begin{tikzpicture}[very thick,scale=0.6,color=blue!50!black, baseline]
\draw[color=green!65!black] (0,-1) -- (0,1); 
\draw[color=green!65!black] (0,-1) node[right] (X) {{\small${}_g I$}};
\fill(-1.2,0) circle (2.9pt) node[above] {{\small$\phi$}};
\end{tikzpicture} 
=
\; 
\begin{tikzpicture}[very thick,scale=0.6,color=blue!50!black, baseline]
\draw[color=green!65!black] (0,-1) -- (0,1); 
\draw[color=green!65!black] (0,-1) node[right] (X) {{\small${}_g I$}};
\fill(1.4,0) circle (2.9pt) node[above] {{\small$g\cdot\phi$}};
\end{tikzpicture} 
$$ 

This brings us to the first step of the reformulation of the orbifold construction: the $g$-twisted sector is identical with defect junction fields $\alpha_g$ between the invisible defect and ${}_g I$: 
$$
H_g = \Hom(I, {}_g I) \ni \alpha_g \equiv 
\begin{tikzpicture}[very thick,scale=0.6,color=green!65!black, baseline]
\fill (0,0) circle (2.9pt) node[right] (D) {{\small $\alpha_g$}};
\fill (0,0.7) circle (0) node[left] (u) {{\small ${}_g I$}};
\fill (0,-0.7) circle (0) node[left] (d) {{\small $I$}};
\draw (0,0) -- (0,0.9); 
\draw[dashed] (0,0) -- (0,-0.9); 
\end{tikzpicture} 
\equiv  
\begin{tikzpicture}[very thick,scale=0.6,color=green!65!black, baseline]
\fill (0,-0.5) circle (2.9pt) node[right] (D) {{\small $\alpha_g$}};
\draw (0,-0.5) -- (0,0.6); 
\end{tikzpicture} 
. 
$$
The second step is also immediate: for any defect its induced action on a field~$\phi$ is given by wrapping it around the insertion point of~$\phi$; hence acting with $h\in G$ on an element of the $g$-twisted sector corresponds to encircling it with ${}_h I$. Thus the orbifold projector~$P$ acting on $\alpha_g \in \Hom(I, {}_g I)$ is
$$
\begin{tikzpicture}[very thick,scale=0.6,color=green!65!black, baseline]
\fill (0,-0.5) circle (2.9pt) node[left] (D) {{\small $\alpha_g$}};
\draw (0,-0.5) -- (0,0.6); 
\end{tikzpicture} 
\lmt 
\frac{1}{|G|}  \sum_{h\in G} h(\alpha_g)
=
\frac{1}{|G|} 
\sum_{h\in G}
\begin{tikzpicture}[very thick,scale=0.75,color=green!65!black, baseline=0.2cm]
\draw (0,-0.3) -- (0,1.3);
\fill (0,-0.3) circle (2.25pt) node[left] {};
\fill (0.1,-0.3) circle (0pt) node[left] {{\small$\alpha_g$}};
\draw (0,0.8) .. controls +(-0.9,-0.3) and +(-0.9,0) .. (0,-0.8);
\draw[
	decoration={markings, mark=at position 0.8 with {\arrow{<}}}, postaction={decorate}
	]
	 (0,-0.8) .. controls +(0.9,0) and +(0.9,0.8) .. (0,0.1);
\draw[->] (0.01,-0.8) -- (-0.01,-0.8);
\fill (0,0.1) circle (2.25pt) node {};
\fill (0,0.8) circle (2.25pt) node {};
\fill (-0.55,-0.05) circle (0pt) node[left] {{\tiny $h\vphantom{h^{-1}}$}};
\fill (-0.07,-0.0) circle (0pt) node[right] {{\tiny $g\vphantom{h^{-1}}$}};
\fill (1.4,-0.05) circle (0pt) node[left] {};
\fill (-0.1,0.65) circle (0pt) node[right] {{\tiny $gh^{-1}$}};
\fill (1.30,1.2) circle (0pt) node[left] {{\tiny $hgh^{-1}$}};
\end{tikzpicture}
$$
where the trivalent junction fields 
$
\!\!\!
\begin{tikzpicture}[very thick,scale=0.25,color=green!65!black, baseline=0.05cm]
\draw[-dot-] (3,0) .. controls +(0,1) and +(0,1) .. (2,0);
\draw (2.5,0.75) -- (2.5,1.5); 
\fill (2,0) circle (0pt) node[left] (D) {};
\fill (3,0) circle (0pt) node[right] (D) {};
\fill (2.5,1.5) circle (0pt) node[right] (D) {};
\end{tikzpicture} 
\!\!\!\!=\mu_{k,l}:{}_k I \otimes {}_l I \rightarrow {}_{kl} I$
and their inverses 
$
\begin{tikzpicture}[very thick,scale=0.25,color=green!65!black, baseline=-0.3cm, rotate=180]
\draw[-dot-] (3,0) .. controls +(0,1) and +(0,1) .. (2,0);
\draw (2.5,0.75) -- (2.5,1.5); 
\fill (2,0) circle (0pt) node[left] (D) {};
\fill (3,0) circle (0pt) node[right] (D) {};
\fill (2.5,1.5) circle (0pt) node[right] (D) {};
\end{tikzpicture} 
\!\!$ 
implement multiplication in~$G$ as fusion between defects. For this the $\mu_{k,l}$ must satisfy an associativity condition with an obstruction in $H^3(G, U(1))$, see e.\,g.~\cite{ffrs0909.5013}. 

In summary, the orbifold bulk space~$\mathcal H$ can be described purely in terms of the fundamental defect
$$
A_G = \bigoplus_{g\in G} {}_g I
$$
which is topological by the assumption that~$G$ is a symmetry of the unorbifolded theory. 
The space $\mathcal H$ is the image of the orbifold projector acting on $H = \Hom(I,A_G) = \bigoplus_{g\in G} \Hom(I,{}_g I)$ as
\be\label{eq:bulk-AG-action}
H \ni \!
\begin{tikzpicture}[very thick,scale=0.6,color=green!50!black, baseline]
\fill (0,-0.5) circle (2.9pt) node[below] (D) {{\small $\alpha$}};
\draw (0,-0.5) -- (0,0.6); 
\fill (0,0.6) circle (0pt) node[above] (D) {{\small $A_G$}};
\end{tikzpicture} 
\lmt 
\begin{tikzpicture}[very thick,scale=0.6,color=green!50!black, baseline=0.2cm]
\draw (0,0) -- (0,1.3);
\fill (0,0) circle (2.9pt) node[below] {{\small$\alpha$}};
\draw (0,0.8) .. controls +(-0.9,-0.3) and +(-0.9,0) .. (0,-0.8);
\draw[
	decoration={markings, mark=at position 0.83 with {\arrow{<}}}, postaction={decorate}
	]
	 (0,-0.8) .. controls +(0.9,0) and +(0.9,0.8) .. (0,0.4);
\draw[->] (0.01,-0.8) -- (-0.01,-0.8);
\fill (0,0.4) circle (2.9pt) node {};
\fill (0.08,0.35) circle (0pt) node[left] (D) {};
\fill (0,0.8) circle (2.9pt) node {};
\fill (-0.08,0.8) circle (0pt) node[right] (D) {};
\fill (0,1.3) circle (0pt) node[left] (D) {{\small $A_G$}};
\end{tikzpicture}
. 
\ee
Here the `(co)multiplication junction fields' $\mu = \!\!\!
\begin{tikzpicture}[very thick,scale=0.25,color=green!50!black, baseline=0.05cm]
\draw[-dot-] (3,0) .. controls +(0,1) and +(0,1) .. (2,0);
\draw (2.5,0.75) -- (2.5,1.5); 
\fill (2,0) circle (0pt) node[left] (D) {};
\fill (3,0) circle (0pt) node[right] (D) {};
\fill (2.5,1.5) circle (0pt) node[right] (D) {};
\end{tikzpicture} 
\!\!\!\!
: A_G \otimes A_G \rightarrow A_G$ and $\Delta = 
\begin{tikzpicture}[very thick,scale=0.25,color=green!50!black, baseline=-0.3cm, rotate=180]
\draw[-dot-] (3,0) .. controls +(0,1) and +(0,1) .. (2,0);
\draw (2.5,0.75) -- (2.5,1.5); 
\fill (2,0) circle (0pt) node[left] (D) {};
\fill (3,0) circle (0pt) node[right] (D) {};
\fill (2.5,1.5) circle (0pt) node[right] (D) {};
\end{tikzpicture} 
\!\!
: A_G \rightarrow A_G \otimes A_G$ are given in terms of the group multiplications $\mu_{g,h}$ as $\sum_{g,h\in G} \mu_{g,h}$ and $\frac{1}{|G|}\sum_{g,h\in G} \mu_{g,h}^{-1}$, respectively. To avoid clutter we do not label the associated trivalent junctions in diagrams like the above. 

\medskip

Next we explain how to describe orbifold boundary and defect sectors in defect language. It turns out that equivariance of a defect~$X$ between two unorbifolded theories with symmetry groups~$G$ and~$G'$ is precisely captured by compatible left action of $A_G$ and right action of $A_{G'}$ on~$X$ via fusion in the following sense. $X$ is a defect in the orbifold theory iff it comes with special junction fields $\rho_{\text{l}}: A_G \otimes X \rightarrow X$ and $\rho_{\text{r}}: X \otimes A_{G'} \rightarrow X$ such that
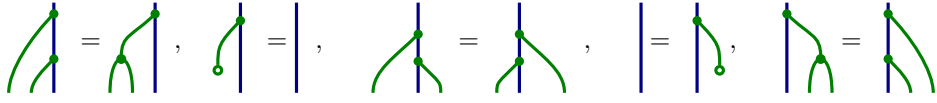
\be\label{eq:bimodule}
\begin{tikzpicture}[very thick,scale=0.60,color=blue!50!black, baseline]
\draw (0,-1) node[right] (X) {};
\draw (0,1) node[right] (Xu) {};
\draw (0,-1) -- (0,1); 
\fill[color=green!50!black] (0,-0.25) circle (2.9pt) node (meet) {};
\fill[color=green!50!black] (0,0.75) circle (2.9pt) node (meet2) {};
\draw[color=green!50!black] (-0.5,-1) .. controls +(0,0.25) and +(-0.25,-0.25) .. (0,-0.25);
\draw[color=green!50!black] (-1,-1) .. controls +(0,0.5) and +(-0.5,-0.5) .. (0,0.75);
\end{tikzpicture} 
=
\begin{tikzpicture}[very thick,scale=0.60,color=blue!50!black, baseline]

\draw (0,-1) node[right] (X) {};
\draw (0,1) node[right] (Xu) {};
\draw (0,-1) -- (0,1); 
\fill[color=green!50!black] (0,0.75) circle (2.9pt) node (meet2) {};
\draw[-dot-, color=green!50!black] (-0.5,-1) .. controls +(0,1) and +(0,1) .. (-1,-1);
\draw[color=green!50!black] (-0.75,-0.2) .. controls +(0,0.5) and +(-0.5,-0.5) .. (0,0.75);
\end{tikzpicture} 
, \quad
\begin{tikzpicture}[very thick,scale=0.60,color=blue!50!black, baseline]
\draw (0,-1) -- (0,1); 
\draw (0,-1) node[right] (X) {};
\draw (0,1) node[right] (Xu) {};
\draw[color=green!50!black]  (-0.5,-0.5) node[Odot] (unit) {}; 
\fill[color=green!50!black]  (0,0.6) circle (2.9pt) node (meet) {};
\draw[color=green!50!black]  (unit) .. controls +(0,0.5) and +(-0.5,-0.5) .. (0,0.6);
\end{tikzpicture} 
=
\begin{tikzpicture}[very thick,scale=0.60,color=blue!50!black, baseline]
\draw (0,-1) -- (0,1); 
\draw (0,-1) node[right] (X) {};
\draw (0,1) node[right] (Xu) {};
\end{tikzpicture} 
, \quad
\begin{tikzpicture}[very thick,scale=0.6,color=blue!50!black, baseline]
\draw (0,-1) node[left] (X) {};
\draw (0,1) node[left] (Xu) {};
\draw (0.5,-1) node[right] (A) {};
\draw (-1,-1) node[left] (B) {};
\draw (0,-1) -- (0,1); 
\fill[color=green!50!black] (0,-0.3) circle (2.9pt) node (meet) {};
\fill[color=green!50!black] (0,0.3) circle (2.9pt) node (meet) {};
\draw[color=green!50!black] (-1,-1) .. controls +(0,0.5) and +(-0.5,-0.5) .. (0,0.3);
\draw[color=green!50!black] (0.5,-1) .. controls +(0,0.25) and +(0.5,-0.5) .. (0,-0.3);
\end{tikzpicture} 
\!\!
=
\!\!
\begin{tikzpicture}[very thick,scale=0.60,color=blue!50!black, baseline]
\draw (0,-1) node[right] (X) {};
\draw (0,1) node[right] (Xu) {};
\draw (1.0,-1) node[right] (A) {};
\draw (-0.5,-1) node[left] (B) {};
\draw (0,-1) -- (0,1); 
\fill[color=green!50!black] (0,-0.3) circle (2.9pt) node (meet) {};
\fill[color=green!50!black] (0,0.3) circle (2.9pt) node (meet) {};
\draw[color=green!50!black] (-0.5,-1) .. controls +(0,0.25) and +(-0.5,-0.5) .. (0,-0.3);
\draw[color=green!50!black] (1.0,-1) .. controls +(0,0.5) and +(0.5,-0.5) .. (0,0.3);
\end{tikzpicture} 
, \quad
\begin{tikzpicture}[very thick,scale=0.60,color=blue!50!black, baseline]
\draw (0,-1) -- (0,1); 
\draw (0,-1) node[left] (X) {};
\draw (0,1) node[left] (Xu) {};
\end{tikzpicture} 
=
\begin{tikzpicture}[very thick,scale=0.60,color=blue!50!black, baseline]
\draw (0,-1) -- (0,1); 
\draw (0,-1) node[left] (X) {};
\draw (0,1) node[left] (Xu) {};
\draw[color=green!50!black]  (0.5,-0.5) node[Odot] (unit) {}; 
\fill[color=green!50!black]  (0,0.6) circle (2.9pt) node (meet) {};
\draw[color=green!50!black]  (unit) .. controls +(0,0.5) and +(0.5,-0.5) .. (0,0.6);
\end{tikzpicture} 
\, , \quad
\begin{tikzpicture}[very thick,scale=0.60,color=blue!50!black, baseline]
\draw (0,-1) node[left] (X) {};
\draw (0,1) node[left] (Xu) {};
\draw (0,-1) -- (0,1); 
\fill[color=green!50!black] (0,0.75) circle (2.9pt) node (meet2) {};
\draw[-dot-, color=green!50!black] (0.5,-1) .. controls +(0,1) and +(0,1) .. (1,-1);
\draw[color=green!50!black] (0.75,-0.2) .. controls +(0,0.5) and +(0.5,-0.5) .. (0,0.75);
\end{tikzpicture} 
= 
\begin{tikzpicture}[very thick,scale=0.60,color=blue!50!black, baseline]
\draw (0,-1) node[left] (X) {};
\draw (0,1) node[left] (Xu) {};
\draw (0,-1) -- (0,1); 
\fill[color=green!50!black] (0,-0.25) circle (2.9pt) node (meet) {};
\fill[color=green!50!black] (0,0.75) circle (2.9pt) node (meet2) {};
\draw[color=green!50!black] (0.5,-1) .. controls +(0,0.25) and +(0.25,-0.25) .. (0,-0.25);
\draw[color=green!50!black] (1,-1) .. controls +(0,0.5) and +(0.5,-0.5) .. (0,0.75);
\end{tikzpicture} 
\ee
where we write 
$
\begin{tikzpicture}[very thick,scale=0.4,color=green!50!black, baseline=-0.2cm]
\draw (0,-0.5) node[Odot] (D) {}; 
\draw (D) -- (0,0.3); 
\end{tikzpicture} 
$
for the embedding of $I = {}_e I$ into both $A_G$ and $A_{G'}$.\footnote{Note that for us generic defects are blue while special defects like $A_G$ or $A_{G'}$ are green.} One says that~$X$ is an $A_G$-$A_{G'}$-bimodule. As noted before, boundary conditions~$Q$ are defects to one side of which lies the trivial theory. In our convention this translates into the map~$\rho_{\text{r}}$ being trivial in this case, making~$Q$ a (left) module over $A_G$. 

Finally, junction fields in the orbifold theory are maps between equivariant defects~$X$ and~$Y$ that commute with the actions of $A_G$ and $A_{G'}$. We denote the space of all such maps $\Hom_{A_G A_{G'}}(X,Y)$, and we write $\Hom_{A_G}(Q,P)$ for boundary operators between~$Q$ and~$P$. A special example of an $A_G$-$A_G$-bimodule is $A_G$ itself, with $\rho_{\text{l}} = \rho_{\text{r}} = \mu$. Thus it is a defect in the orbifold theory, whose bulk space is equivalently described as $\End_{A_G A_G}(A_G)$, made up of fields living on $A_G$ that commute with~$\mu$. In other words, $A_G$ is the invisible defect in the orbifold theory. 

\medskip

The question arises whether the above description of orbifolds is more than a mere exercise in defect prose. It is, for at least two reasons: 
the universality of the defect perspective provides conceptual clarity which leads directly to the generalisations to be touched upon in the next section, and moreover gives a framework in which concrete computations can be carried out very efficiently. 
Both of these aspects are rooted in the following 

\begin{punchline}
Every physical picture of defects and field insertions on the worldsheet, such as~\eqref{eq:bulk-AG-action}, translates directly into a computable, rigorous mathematical expression. 
\end{punchline}

As explained in \cite{dkr1107.0495, cm1208.1481, genorb, BCP} the natural setting is that of `pivotal bicategories', but we do not want to stress this here. All we need to know in the following is that locally every physical diagram translates into a single composite defect junction field by reading it from bottom to top (operator product) and from right to left (fusion product): 
$$
\begin{tikzpicture}[thick,scale=0.6,color=blue!50!black, baseline=0cm]
\draw[line width=0] 
(0,1.25) node[line width=0pt] (A) {{\small $Z$}}
(0,-1.25) node[line width=0pt] (A2) {{\small $X$}};
\draw (A2) -- (A); 
\fill (0,0.4) circle (3.3pt) node[left] {{\small$\psi$}};
\fill (0,-0.4) circle (3.3pt) node[left] {{\small$\phi$}};
\end{tikzpicture}
= 
\begin{tikzpicture}[thick,scale=0.6,color=blue!50!black, baseline=0cm]
\draw[line width=0] 
(0,1.25) node[line width=0pt] (A) {{\small $Z$}}
(0,-1.25) node[line width=0pt] (A2) {{\small $X$}};
\draw (A2) -- (A); 
\fill (0,0) circle (3.3pt) node[left] {{\small$\psi\phi$}};
\end{tikzpicture}
\, , \quad
\begin{tikzpicture}[thick,scale=0.6,color=blue!50!black, baseline=0cm]
\draw[line width=0] 
(0,1.25) node[line width=0pt] (A) {{\small $Y\vphantom{X'}$}}
(0,-1.25) node[line width=0pt] (A2) {{\small $X\vphantom{X'}$}};
\draw (A2) -- (A); 
\draw[line width=0] 
(0.7,1.25) node[line width=0pt] (B) {{\small $Y'$}}
(0.7,-1.25) node[line width=0pt] (B2) {{\small $X'$}};
\draw (B2) -- (B); 
\fill (0,0) circle (3.3pt) node[left] {{\small$\varphi\vphantom{\varphi'}$}};
\fill (0.7,0) circle (3.3pt) node[right] {{\small$\varphi'$}};
\end{tikzpicture}
= 
\begin{tikzpicture}[thick,scale=0.6,color=blue!50!black, baseline=0cm]
\draw[line width=0] 
(0,1.25) node[line width=0pt] (A) {{\small $Y\!\otimes\! Y'$}}
(0,-1.25) node[line width=0pt] (A2) {{\small $XÊ\!\otimes\! X'$}};
\draw[ultra thick] (A2) -- (A); 
\fill (0,0) circle (3.3pt) node[left] {{\small$\varphi \!\otimes\! \varphi' $}};
\end{tikzpicture}
.
$$
Apart from naturally occurring junction fields such as $\alpha, \mu, \rho_{\text{l}}, \rho_{\text{r}}$ above we have the following universal fields for any defect~$X$: isomorphisms $\lambda_X: IÊ\otimes X \cong X$, $\rho_X: X \otimes I \cong X$ implementing the trivial fusion with~$I$, and (co)evaluation maps
\begin{align}
\label{eq:adjunctionmaps}
\ev_X = & 
\begin{tikzpicture}[very thick,scale=0.65,color=blue!50!black, baseline=.25cm]
\draw[line width=0pt] 
(3,0) node[line width=0pt] (D) {}
(2,0) node[line width=0pt] (s) {}; 
\draw[directed] (D) .. controls +(0,1) and +(0,1) .. (s);
\end{tikzpicture}
\!\! : X^\dagger \otimes X 
\lra I
\, , \quad
\coev_X = 
\begin{tikzpicture}[very thick,scale=0.65,color=blue!50!black, baseline=-.4cm,rotate=180]
\draw[line width=0pt] 
(3,0) node[line width=0pt] (D) {}
(2,0) node[line width=0pt] (s) {}; 
\draw[redirected] (D) .. controls +(0,1) and +(0,1) .. (s);
\end{tikzpicture}
\!\! : I \lra X \otimes X^\dagger
\, , \\
\tev_X = & 
\begin{tikzpicture}[very thick,scale=0.65,color=blue!50!black, baseline=.25cm]
\draw[line width=0pt] 
(3,0) node[line width=0pt] (D) {}
(2,0) node[line width=0pt] (s) {}; 
\draw[redirected] (D) .. controls +(0,1) and +(0,1) .. (s);
\end{tikzpicture}
\!\! : X \otimes X^\dagger 
\lra I
\, , \quad 
\tcoev_X = 
\begin{tikzpicture}[very thick,scale=0.65,color=blue!50!black, baseline=-.4cm,rotate=180]
\draw[line width=0pt] 
(3,0) node[line width=0pt] (D) {}
(2,0) node[line width=0pt] (s) {}; 
\draw[directed] (D) .. controls +(0,1) and +(0,1) .. (s);
\end{tikzpicture}
\!\! : I \lra X^\dagger \otimes X
\end{align}
exhibiting the orientation-reversed defect $X^\dagger$ as the left and right adjoint of~$X$.\footnote{In general left and right adjoints do not have to be equal. In the present note we gloss over this and refer to the above references for a detailed treatment of this issue.} 

This list exhausts all the special junction fields we will ever need. For example, the map on the right-hand side of~\eqref{eq:bulk-AG-action} from~$I$ to $A_G$ is a concatenation of 
$\coev_{A_G}$, 
$\alpha$, 
$\Delta$, 
$\tev_{A_G}$, 
and $\mu$, in that order. 
In concrete examples such as two-dimensional Yang-Mills theory or Landau-Ginzburg models all these maps are explicitly known. 

\medskip

Of special interest are two-dimensional $\mathcal N=(2,2)$ superconformal field theories and their topological twists. Even if their Neveu-Schwarz and Ramond sectors are isomorphic, i.\,e.~if spectral flow corresponds to a state in the theory, this does not have to be true in the orbifold theory \cite{IntriligatorVafa1990}. How then can the defect description encompass both sectors? Our answer is that there are actually two natural ways to wrap the symmetry defect $A_G$ around twisted sector fields, thus providing two possibly distinct orbifold projections. One is~\eqref{eq:bulk-AG-action} above, and we claim that this constructs the bulk space of RR ground states, which is isomorphic to 
\be\label{eq:HRRspace}
\mathcal{H}_{\text{RR}} = 
\im \Big\{ 
\Hom(I,A_G) \ni \!\!
\begin{tikzpicture}[very thick,scale=0.6,color=green!50!black, baseline]
\fill (0,-0.5) circle (2.9pt) node[left] (D) {};
\draw (0,-0.5) -- (0,0.6); 
\end{tikzpicture} 
\lmt 
\begin{tikzpicture}[very thick,scale=0.6,color=green!50!black, baseline=0.2cm]
\draw (0,0) -- (0,1.3);
\fill (0,0) circle (2.9pt) node[left] {};
\draw (0,0.8) .. controls +(-0.9,-0.3) and +(-0.9,0) .. (0,-0.8);
\draw[
	decoration={markings, mark=at position 0.83 with {\arrow{<}}}, postaction={decorate}
	]
	 (0,-0.8) .. controls +(0.9,0) and +(0.9,0.8) .. (0,0.4);
\draw[->] (0.01,-0.8) -- (-0.01,-0.8);
\fill (0,0.4) circle (2.9pt) node {};
\fill (0,0.8) circle (2.9pt) node {};
\end{tikzpicture}
\Big\} 
\, .
\ee
On the other hand, $A_G$ can also be wrapped without enlisting the adjunction maps, 
and this leads to the space of (c,c) fields: 
$$
\mathcal{H}_{\text{(c,c)}} = 
\im \Big\{ 
\Hom(I,A_G) \ni \!\!
\begin{tikzpicture}[very thick,scale=0.6,color=green!50!black, baseline]
\fill (0,-0.5) circle (2.9pt) node[left] (D) {};
\draw (0,-0.5) -- (0,0.6); 
\end{tikzpicture} 
\lmt 
\begin{tikzpicture}[very thick,scale=0.6,color=green!50!black, baseline]
\draw (0,0) -- (0,1.25);
\fill (0,0) circle (2.9pt) node[left] {};
\draw (0,0.8) .. controls +(-0.9,-0.3) and +(-0.9,0) .. (0,-0.8);
\draw (0,-0.8) .. controls +(0.9,0) and +(0.7,-0.1) .. (0,0.4);
\fill (0,-0.8) circle (2.9pt) node {};
\fill (0,0.4) circle (2.9pt) node {};
\fill (0,0.8) circle (2.9pt) node {};
\draw (0,-1.2) node[Odot] (unit) {};
\draw (0,-0.8) -- (unit);
\end{tikzpicture}
\Big\} 
\, .
$$

As explained in \cite{BCP} the projectors to $\mathcal{H}_{\text{RR}}$ and $\mathcal{H}_{\text{(c,c)}}$ are related by
$$
\begin{tikzpicture}[very thick,scale=0.6,color=green!50!black, baseline=0.2cm]
\draw (0,0) -- (0,1.3);
\fill (0,0) circle (2.9pt) node[left] {};
\draw (0,0.8) .. controls +(-0.9,-0.3) and +(-0.9,0) .. (0,-0.8);
\draw[
	decoration={markings, mark=at position 0.83 with {\arrow{<}}}, postaction={decorate}
	]
	 (0,-0.8) .. controls +(0.9,0) and +(0.9,0.8) .. (0,0.4);
\draw[->] (0.01,-0.8) -- (-0.01,-0.8);
\fill (0,0.4) circle (2.9pt) node {};
\fill (0,0.8) circle (2.9pt) node {};
\end{tikzpicture}
= 
\begin{tikzpicture}[very thick,scale=0.6,color=green!50!black, baseline=0.1cm]
\draw (0,0) -- (0,1.3);
\fill (0,0) circle (2.9pt) node[left] {};
\fill (-0.67,0) circle (2.9pt) node[left] {{\small $\gamma_{A_G}$}};
\draw (0,0.8) .. controls +(-0.9,-0.3) and +(-0.9,0) .. (0,-0.8);
\draw (0,-0.8) .. controls +(0.9,0) and +(0.7,-0.1) .. (0,0.4);
\fill (0,-0.8) circle (2.9pt) node {};
\fill (0,0.4) circle (2.9pt) node {};
\fill (0,0.8) circle (2.9pt) node {};
\draw (0,-1.2) node[Odot] (unit) {};
\draw (0,-0.8) -- (unit);
\end{tikzpicture}
\, , \qquad \text{where} \quad
\gamma_{A_G} = 
\begin{tikzpicture}[very thick, scale=0.45,color=green!50!black, baseline=-0.35cm]
\draw (0,0.8) -- (0,2);
\draw[-dot-] (0,0.8) .. controls +(0,-0.5) and +(0,-0.5) .. (-0.75,0.8);
\draw[directedgreen, color=green!50!black] (-0.75,0.8) .. controls +(0,0.5) and +(0,0.5) .. (-1.5,0.8);
\draw[-dot-] (0,-1.8) .. controls +(0,0.5) and +(0,0.5) .. (-0.75,-1.8);
\draw[redirectedgreen, color=green!50!black] (-0.75,-1.8) .. controls +(0,-0.5) and +(0,-0.5) .. (-1.5,-1.8);
\draw (0,-1.8) -- (0,-3);
\draw (-1.5,0.8) -- (-1.5,-1.8);
\draw (-0.375,-0.2) node[Odot] (D) {}; 
\draw (-0.375,0.4) -- (D);
\draw (-0.375,-0.8) node[Odot] (E) {}; 
\draw (-0.375,-1.4) -- (E);
\end{tikzpicture}
$$
is the so-called \textsl{Nakayama automorphism} of $A_G$. Hence the two bulk spaces coincide if $\gamma_{A_G} = 1_{A_G}$. We note in passing that $\mathcal{H}_{\text{RR}}$ and $\mathcal{H}_{\text{(c,c)}}$ can also be interpreted as Hochschild homology and cohomology of $A_G$. 

There are similar NS and R projectors in the boundary and defect sectors. More precisely, boundary operators $Q\rightarrow P$ and defect junction fields $X\rightarrow Y$ form the spaces $\Hom_{A_G}(Q,P)$, $\Hom_{A_G A_{G'}}(X,Y)$ in the NS sector and $\Hom_{A_G}(Q,{}_{\gamma_{A_G}} P)$, $\Hom_{A_G A_{G'}}(X,{}_{\gamma_{A_G}} Y)$ in the R sector. Here ${}_{\gamma_{A_G}}(-)$ denotes the twist of the left module action~$\rho_{\text{l}}$ on, say, $Y$ by the Nakayama automorphism, i.\,e.~the map $A_G \otimes Y \rightarrow Y$ is now $\rho_{\text{l}} \circ (\gamma_{A_G} \otimes 1_Y)$. 

\medskip

Finally, a typical example of translating a physical picture into a precise expression is a disc correlator with a twisted bulk field $\alpha \in \Hom(I, A_G)$ and a boundary operator $\Phi: Q\rightarrow Q$. Viewing the boundary as a defect from the trivial theory~$\C$ to some other theory the correlator is
\be\label{eq:disccorrelator}
\begin{tikzpicture}[very thick,scale=0.6,color=blue!50!black, baseline]
\nicepalecolourscheme (0,0) circle (1.5);
%
\draw (0,0) circle (1.5);
\fill (-45:1.55) circle (0pt) node[right] {{\small$Q$}};
\draw[<-, very thick] (0.100,-1.5) -- (-0.101,-1.5) node[above] {}; 
\draw[<-, very thick] (-0.100,1.5) -- (0.101,1.5) node[below] {}; 
\fill[color=green!50!black] (135:0) circle (2.9pt) node[left] {{\small$\alpha$}};
\fill (0:1.5) circle (2.9pt) node[left] {{\small$\Phi$}};
\draw[color=green!50!black] (0,0) .. controls +(0,0.6) and +(-0.4,-0.4) .. (45:1.5);
\fill[color=green!50!black] (45:1.5) circle (2.9pt) node[right] {{\small$\rho_{\text{l}}$}};
\end{tikzpicture} 
\ee
which as a linear map from~$\C$ to~$\C$ is equal to the number 
$
\ev_Q 
\circ
(1\otimes \rho_{\text{l}})
\circ
(1\otimes \alpha \otimes \Phi)
\circ
(1\otimes \lambda_Q^{-1})
\circ
\tcoev_Q
$. 
In particular, if~$\Phi$ is the identity and~$\alpha$ an RR ground state, this diagram computes the RR charge of the brane~$Q$ in the orbifold theory. 

\begin{example}[`Discrete' two-dimensional Yang-Mills theory]
\textup{The simplest unorbifolded theory to start from is the trivial one with bulk space $H_e = \C$. Defects then are $\C$-vector spaces, in particular the identity defect is $I = \C$. Any finite group~$G$ may be regarded as a symmetry with trivial action on~$H_e$. A straightforward computation shows that the defect $A_G = \bigoplus_{g\in G} {}_g I$ is nothing but the group algebra $\C G$ whose multiplication~$\mu$ is the convolution product and whose Nakayama automorphism is the identity. Endomorphisms of $\C G$ viewed as a bimodule over itself are precisely the class functions of~$G$, with a standard basis given by characters of irreducible $G$-representations. Thus we obtain the bulk space of two-dimensional topological Yang-Mills theory for the case that the gauge group is finite \cite{cmr9411210, BrunnerDiploma, pr1996}. Modules over $A_G = \C G$ are in one-to-one correspondence with $G$-representations, and this is precisely the data needed to label Wilson lines. They also occur as defects since every $\C G$-module can be made into a bimodule in a way that is compatible with fusion: the functor $(-)\otimes_\C \C G$ is monoidal.  
} 

\textup{We have thus produced `discrete' two-dimensional Yang-Mills theory in the topological zero-area limit as an orbifold of the trivial theory. To produce `proper' two-dimensional topological Yang-Mills theory with an (infinite) Lie group one may either take a leap of faith and boldly replace sums by integrals etc.~ in the above analysis. Or one may attempt to rigorously extend the orbifold completion construction of \cite{genorb} to infinite groups by formulating it in terms of simplicial objects in the bicategorical setting \cite{CarMur}. 
Orbifolds by compact Lie groups have recently also been studied in \cite{gs1112.1708, fr1208.6136}.}
\end{example}

\begin{example}[Landau-Ginzburg orbifolds]
\textup{
B-twisted affine Landau-Ginzburg models are under very good control with matrix factorisations describing boundary and defect conditions; see \cite{cm1303.1389} for a short review of the relevant properties that includes explicit formulas for the left and right actions $\lambda_X^{\pm 1}, \rho_X^{\pm 1}$ of the identity defect and the adjunction maps~\eqref{eq:adjunctionmaps}. As explained in detail in \cite{genorb, BCP} the general defect construction building on $A_G$ discussed above precisely recovers the conventional orbifold results of the bulk \cite{IntriligatorVafa1990}, boundary \cite{add0401}, and defect \cite{br0712.0188, genorb} sectors. The orbifold spaces $\mathcal{H}_{\text{RR}}$ and $\mathcal{H}_{\text{(c,c)}}$ are often not isomorphic since the Nakayama automorphism is given by
$$
\gamma_{A_G} = \sum_{g\in G} \det(g)^{-1} \cdot 1_{{}_g I}
$$
where $\det(g)$ denotes the determinant of the matrix representing the action of~$g$ on the variables of the unorbifolded Landau-Ginzburg model. Furthermore, any correlator in the orbifold theory can be straightforwardly computed using the rules collected in \cite{cm1303.1389}, cf.~\cite[Sect.\,2.3]{BCP}. This in particular includes the disc correlator~\eqref{eq:disccorrelator}, which allows the proposal of \cite{w0412274} for RR brane charges to be derived as a simple residue formula from first principles. A special case of this formula was recently obtained in \cite{hr1308.2438} as part of an extensive study of hemisphere partition functions and central charges. 
}
\end{example}

\section{Generalised orbifolds via defects}

It is easy to check that in any TFT and for any finite symmetry group~$G$ the junction fields $\mu, \Delta, 
\begin{tikzpicture}[very thick,scale=0.4,color=green!50!black, baseline=-0.2cm]
\draw (0,-0.5) node[Odot] (D) {}; 
\draw (D) -- (0,0.3); 
\end{tikzpicture} 
$ 
and 
$
\begin{tikzpicture}[very thick,scale=0.4,color=green!50!black, baseline=-0.05cm, rotate=180]
\draw (0,-0.5) node[Odot] (D) {}; 
\draw (D) -- (0,0.3); 
\end{tikzpicture} 
: A_G \twoheadrightarrow I
$ 
of the defect $A_G$ satisfy not only the defining (co)associativity and (co)unital conditions
\begin{align}\label{eq:sFA1}
& 
\begin{tikzpicture}[very thick,scale=0.60,color=green!50!black, baseline=0.6cm]
\draw[-dot-] (3,0) .. controls +(0,1) and +(0,1) .. (2,0);
\draw[-dot-] (2.5,0.75) .. controls +(0,1) and +(0,1) .. (3.5,0.75);
\draw (3.5,0.75) -- (3.5,0); 
\draw (3,1.5) -- (3,2.25); 
\end{tikzpicture} 
=
\begin{tikzpicture}[very thick,scale=0.60,color=green!50!black, baseline=0.6cm]
\draw[-dot-] (3,0) .. controls +(0,1) and +(0,1) .. (2,0);
\draw[-dot-] (2.5,0.75) .. controls +(0,1) and +(0,1) .. (1.5,0.75);
\draw (1.5,0.75) -- (1.5,0); 
\draw (2,1.5) -- (2,2.25); 
\end{tikzpicture} 
\, , \quad
\begin{tikzpicture}[very thick,scale=0.60,color=green!50!black, baseline=-0.2cm]
\draw (-0.5,-0.5) node[Odot] (unit) {}; 
\fill (0,0.6) circle (2.9pt) node (meet) {};
\draw (unit) .. controls +(0,0.5) and +(-0.5,-0.5) .. (0,0.6);
\draw (0,-1) -- (0,1); 
\end{tikzpicture} 
=
\begin{tikzpicture}[very thick,scale=0.60,color=green!50!black, baseline=-0.2cm]
\draw (0,-1) -- (0,1); 
\end{tikzpicture} 
=
\begin{tikzpicture}[very thick,scale=0.60,color=green!50!black, baseline=-0.2cm]
\draw (0.5,-0.5) node[Odot] (unit) {}; 
\fill (0,0.6) circle (2.9pt) node (meet) {};
\draw (unit) .. controls +(0,0.5) and +(0.5,-0.5) .. (0,0.6);
\draw (0,-1) -- (0,1); 
\end{tikzpicture} 
\, , \quad
\begin{tikzpicture}[very thick,scale=0.60,color=green!50!black, baseline=-0.75cm, rotate=180]
\draw[-dot-] (3,0) .. controls +(0,1) and +(0,1) .. (2,0);
\draw[-dot-] (2.5,0.75) .. controls +(0,1) and +(0,1) .. (1.5,0.75);
\draw (1.5,0.75) -- (1.5,0); 
\draw (2,1.5) -- (2,2.25); 
\end{tikzpicture} 
=
\begin{tikzpicture}[very thick,scale=0.60,color=green!50!black, baseline=-0.75cm, rotate=180]
\draw[-dot-] (3,0) .. controls +(0,1) and +(0,1) .. (2,0);
\draw[-dot-] (2.5,0.75) .. controls +(0,1) and +(0,1) .. (3.5,0.75);
\draw (3.5,0.75) -- (3.5,0); 
\draw (3,1.5) -- (3,2.25); 
\end{tikzpicture} 
\, , \quad
\begin{tikzpicture}[very thick,scale=0.60,color=green!50!black, baseline=-0.1cm, rotate=180]
\draw (0.5,-0.5) node[Odot] (unit) {}; 
\fill (0,0.6) circle (2.9pt) node (meet) {};
\draw (unit) .. controls +(0,0.5) and +(0.5,-0.5) .. (0,0.6);
\draw (0,-1) -- (0,1); 
\end{tikzpicture} 
=
\begin{tikzpicture}[very thick,scale=0.60,color=green!50!black, baseline=-0.1cm, rotate=180]
\draw (0,-1) -- (0,1); 
\end{tikzpicture} 
=
\begin{tikzpicture}[very thick,scale=0.60,color=green!50!black, baseline=-0.1cm, rotate=180]
\draw (-0.5,-0.5) node[Odot] (unit) {}; 
\fill (0,0.6) circle (2.9pt) node (meet) {};
\draw (unit) .. controls +(0,0.5) and +(-0.5,-0.5) .. (0,0.6);
\draw (0,-1) -- (0,1); 
\end{tikzpicture} 
\end{align}
but also the following identities: 
\begin{align}
&
\label{eq:sFA2}
\begin{tikzpicture}[very thick,scale=0.60,color=green!50!black, baseline=0cm]
\draw[-dot-] (0,0) .. controls +(0,-1) and +(0,-1) .. (-1,0);
\draw[-dot-] (1,0) .. controls +(0,1) and +(0,1) .. (0,0);
\draw (-1,0) -- (-1,1.5); 
\draw (1,0) -- (1,-1.5); 
\draw (0.5,0.8) -- (0.5,1.5); 
\draw (-0.5,-0.8) -- (-0.5,-1.5); 
\end{tikzpicture}
=
\begin{tikzpicture}[very thick,scale=0.60,color=green!50!black, baseline=0cm]
\draw[-dot-] (0,1.5) .. controls +(0,-1) and +(0,-1) .. (1,1.5);
\draw[-dot-] (0,-1.5) .. controls +(0,1) and +(0,1) .. (1,-1.5);
\draw (0.5,-0.8) -- (0.5,0.8); 
\end{tikzpicture}
=
\begin{tikzpicture}[very thick,scale=0.60,color=green!50!black, baseline=0cm]
\draw[-dot-] (0,0) .. controls +(0,1) and +(0,1) .. (-1,0);
\draw[-dot-] (1,0) .. controls +(0,-1) and +(0,-1) .. (0,0);
\draw (-1,0) -- (-1,-1.5); 
\draw (1,0) -- (1,1.5); 
\draw (0.5,-0.8) -- (0.5,-1.5); 
\draw (-0.5,0.8) -- (-0.5,1.5); 
\end{tikzpicture}
\, , \quad 
\begin{tikzpicture}[very thick,scale=0.60,color=green!50!black, baseline=0cm]
\draw[-dot-] (0,0) .. controls +(0,-1) and +(0,-1) .. (1,0);
\draw[-dot-] (0,0) .. controls +(0,1) and +(0,1) .. (1,0);
\draw (0.5,-0.8) -- (0.5,-1.2); 
\draw (0.5,0.8) -- (0.5,1.2); 
\end{tikzpicture}
= 
\begin{tikzpicture}[very thick,scale=0.60,color=green!50!black, baseline=0cm]
\draw (0.5,-1.2) -- (0.5,1.2); 
\end{tikzpicture}
\, . 
\end{align}
This gives $A_G$ the structure of a \textsl{separable Frobenius algebra}. 
Furthermore, depending on~$G$ and the theory at hand, $A_G$ may or may not be \textsl{symmetric}, which is the property 
\be\label{eq:symmetric}
\begin{tikzpicture}[very thick,scale=0.55,color=green!50!black, baseline=0cm]
\draw[-dot-] (0,0) .. controls +(0,1) and +(0,1) .. (-1,0);
\draw[directedgreen, color=green!50!black] (1,0) .. controls +(0,-1) and +(0,-1) .. (0,0);
\draw (-1,0) -- (-1,-1.5); 
\draw (1,0) -- (1,1.5); 
\draw (-0.5,1.2) node[Odot] (end) {}; 
\draw (-0.5,0.8) -- (end); 
\end{tikzpicture}
= 
\begin{tikzpicture}[very thick,scale=0.55,color=green!50!black, baseline=0cm]
\draw[redirectedgreen, color=green!50!black] (0,0) .. controls +(0,-1) and +(0,-1) .. (-1,0);
\draw[-dot-] (1,0) .. controls +(0,1) and +(0,1) .. (0,0);
\draw (-1,0) -- (-1,1.5); 
\draw (1,0) -- (1,-1.5); 
\draw (0.5,1.2) node[Odot] (end) {}; 
\draw (0.5,0.8) -- (end); 
\end{tikzpicture}
\, . 
\ee

It is no accident that these diagrams bear resemblance to the moves
$$
\begin{tikzpicture}[very thick, scale=0.45,color=green!50!black, baseline]
\draw (70:2) -- (0,1);
\draw (110:2) -- (0,1);
\draw (-70:2) -- (0,-1);
\draw (-110:2) -- (0,-1);
\draw (0,1) -- (0,-1);
\end{tikzpicture}
\longleftrightarrow
\begin{tikzpicture}[very thick, scale=0.45,color=green!50!black, baseline]
\draw (150:2) -- (-1,0);
\draw (-150:2) -- (-1,0);
\draw (-30:2) -- (1,0);
\draw (30:2) -- (1,0);
\draw (-1,0) -- (1,0);
\end{tikzpicture}
\, , \quad 
\begin{tikzpicture}[very thick, scale=0.45,color=green!50!black, baseline]
\draw (0,0.8) -- (90:2);
\draw (-1,0) .. controls +(0,1) and +(0,1) .. (1,0);
\draw (-1,0) .. controls +(0,-1) and +(0,-1) .. (1,0);
\draw (0,-0.8) -- (-90:2);
\end{tikzpicture}
\longleftrightarrow
\begin{tikzpicture}[very thick, scale=0.45,color=green!50!black, baseline]
\draw (90:2) -- (-90:2);
\end{tikzpicture}
$$
which allow to transition between any two given triangulations of a surface. The idea is that correlators of an orbifold theory can be computed in the unorbifolded theory by covering the worldsheet with a sufficiently fine network (or triangulation) of the defect $A_G$, thus implementing twisted sectors and orbifold projection. The symmetric separable Frobenius structure ensures that the value of the orbifold correlator is independent of the choice of triangulation. 

\begin{punchline}\cite{ffrs0909.5013, genorb}
The orbifold procedure of Section~\ref{sec:orbifoldsviadefects} via the defect $A_G$ only depends on the fact that $A_G$ has the structure of a (symmetric) separable Frobenius algebra. The construction works in exactly the same way for \textsl{any} defect~$A$ that comes with junction fields such that~\eqref{eq:sFA1}, \eqref{eq:sFA2} and~\eqref{eq:symmetric} hold. In this case~$A$ should be thought of as a `generalised\footnote{Regrettably though ``everyone calls everything generalised something'' and ``a year later you are embarrassed by what you called generalised before'' \cite{GannonTalk}.} gauge symmetry', replacing the orbifold group~$G$, and we call the resulting theory a `generalised orbifold'.\footnote{Dropping the symmetry condition~\eqref{eq:symmetric} leads to TFTs defined on spin or framed surfaces, cf.~\cite[Thm.\,4.19]{TelemanLectures}, \cite[Rem.\,3.9\,\&\,3.10]{BCP} and \cite{NovakRunkel} for details.}
\end{punchline}

Literally everything that we have discussed so far (and more, see \cite{BCP}) also works for generalised orbifolds: starting from a given bulk theory, denoted simply `$a$', together with a defect~$A$ living in it that is a (symmetric) separable Frobenius algebra, we obtain the generalised orbifold denoted $(a,A)$. Its invisible defect is~$A$, its `(c,c)' bulk space is $\End_{AA}(A)$, its boundary conditions $Q,P,\ldots$ are those of~$a$ which have the structure of a left $A$-module (see~\eqref{eq:bimodule}), and boundary operators $Q\rightarrow P$ are elements of $\Hom_A(Q,P)$. Furthermore, defects between $(a,A)$ and another generalised orbifold $(b,B)$ are defects $X, Y,\ldots$ between~$a$ and~$b$ that are also $B$-$A$-bimodules as in~\eqref{eq:bimodule}, and defect junction fields constitute $\Hom_{BA}(X,Y)$. 

One way to construct generalised orbifolds from defects which are not of the form $A_G$ for some group~$G$ is as follows \cite[Thm.\,4.3]{genorb}. Let $X:a\rightarrow b$ be any defect between two possibly distinct (unorbifolded) theories. Then the fusion $A = X^\dagger \otimes X: a\rightarrow a$ is always symmetric and Frobenius, and if the \textsl{quantum dimension} $\dim(X) = \tev_X \circ \coev_X \in \End(I_a)$ is invertible, then~$A$ is also separable. 
One finds that the orbifold $(a,A)$ equivalently describes the theory~$b$. In the bulk this means that the $\C$-algebra $\End_{AA}(A)$ is isomorphic to the bulk space $\End(I_b)$, while the boundary and defect categories of~$b$ are equivalent to the categories of modules and bimodules over~$A$, respectively. 
For instance, the orbifold equivalences of Example~\ref{ex:ADELG} arise in this way, and in the case of E-type singularities~$A$ is not of the form $A_G$, thus giving truly \textsl{generalised} orbifolds. 

\begin{example}[Rational conformal field theory]
\textup{
Generalised orbifolds work in any two-dimensional quantum field theory. The pioneering work started in \cite{tft1} is in the framework of rational CFT, see also \cite{ffrs0909.5013} for a short overview. One important result is roughly that any two consistent rational CFTs of fixed central charge and identical left and right chiral symmetries are generalised orbifolds of one another. This is in particular true for minimal CFTs with and without supersymmetry which come in an ADE classification. 
}
\end{example}

\begin{example}[ADE Landau-Ginzburg models]\label{ex:ADELG}
\textup{
Motivated by the CFT/LG correspondence one expects similar results for Landau-Ginzburg models. Indeed, in \cite[Sect.\,7.3]{genorb} a defect~$A_d$ of the A-type model with potential $u^{2d}$ (concretely a 4-by-4 matrix factorisation of $u^{2d} - v^{2d}$) was constructed such that the D-type model with potential $W_D = x^d-xy^2+z^2$ is a (generalised) $A_d$-orbifold. This is an equivalence not only in the bulk (known since the 80s, see e.\,g.~\cite{ls1990}), but also in the boundary and defect sectors. Concretely, the bulk equivalence reads $\End_{A_d A_d}(A_d) \cong \C[x,y,z]/(\partial W_D)$, while in the open sector we have an equivalence of categories $\hmf(\C[x,y,z], W_D) \cong \operatorname{mod}(A_d)$ between matrix factorisations of $W_D$ and modules over $A_d$, and a similar relation holds in the defect sector. 
}

\textup{
Guided by the corresponding results in rational CFT \cite{o0111139, tft1}, the E-type Landau-Ginzburg models with potentials $x^3 + y^4 + z^2$, $x^3 + xy^3 + z^2$, $x^3 + y^5 + z^2$ should be generalised orbifolds of the defects 
\begin{align*}
& I \oplus P_{\{-3,-2,\ldots,3\}}
\, , \quad
I \oplus P_{\{-4,-3,\ldots,4\}} \oplus P_{\{-8,-7,\ldots,8\}}
\, , \\Ê
& I \oplus P_{\{-5,-4,\ldots,5\}} \oplus P_{\{-9,-8,\ldots,9\}} \oplus P_{\{-14,-13,\ldots,14\}}
\end{align*}
of the A-type models with potential $u^{12}$, $u^{18}$, $u^{30}$, respectively, where we use the notation $P_S$ for `permutation defects' of \cite{br0707.0922}. Just as in the D-type case this has now the status of a theorem \cite{CRCR}. 
}
\end{example}

We conclude with a brief discussion of some of the general features of the construction under consideration here, starting with the bulk and boundary sector. 
One can show \cite{genorb, BCP} that generalised orbifolds satisfy all the axioms of two-dimensional open/closed TFT as described in \cite{l0010269, ms0609042}. If one drops the symmetry condition~\eqref{eq:symmetric} on the orbifolding defect~$A$ and only asks for it to be a separable Frobenius algebra, this is no longer the case, but the axioms are `shared' between the `generalised R and NS sectors' in an interesting way. We present the results in the following, refering to \cite{BCP} for all further details. 

First of all, the operator product and sphere correlator for the orbifold bulk spaces are defined as
$$
\alpha_1 \cdot \alpha_2 = 
\begin{tikzpicture}[very thick,scale=0.75,color=green!50!black, baseline=0.4cm]
\draw[-dot-] (3,0) .. controls +(0,1) and +(0,1) .. (2,0);
\draw (2.5,0.75) -- (2.5,1.5); 
\fill (2,0) circle (2.5pt) node[left] (alpha) {{\small $\alpha_1$}};
\fill (3,0) circle (2.5pt) node[right] (beta) {{\small $\alpha_2$}};
\end{tikzpicture} 
\, , \quad 
\big\langle\alpha_1,\alpha_2 \big\rangle_{(a,A)} 
= \left\langle
\begin{tikzpicture}[very thick,scale=0.75,color=green!50!black, baseline=0.4cm]
\draw[-dot-] (3,0) .. controls +(0,1) and +(0,1) .. (2,0);
\fill (2,0) circle (2.5pt) node[left] (alpha) {{\small $\alpha_1$}};
\fill (3,0) circle (2.5pt) node[right] (beta) {{\small $\alpha_2$}};
\draw (2.5,1.3) node[Odot] (unit) {};
\draw (2.5,0.75) -- (unit); 
\end{tikzpicture} 
\right\rangle_{\raisemath{4pt}{\!\!\!a}} \, .
$$
The product makes $\Hcc \cong \End_{AA}(A)$ into a commutative $\C$-algebra, and $\Hrr \cong \Hom_{AA}(A,{}_{\gamma_A} A)$ is a module over this algebra. The pairing $\langle -,- \rangle _{(a,A)}$ is nondegenerate when restricted to $\Hrr$. 

For any $A$-module~$Q$ the bulk-boundary map $\beta_Q$ and boundary-bulk map $\beta^Q$ are given by
$$
\beta_Q(\alpha) = 
\begin{tikzpicture}[very thick,scale=0.7,color=blue!50!black, baseline]
\nicepalecolourscheme (-1.25,-1.25) rectangle (0,1.25);
\draw (0,-1) -- (0,1); 
\draw (0,-1) node[right] (X) {{\small$Q$}};
\draw (0,1) node[right] (Xu) {{\small$Q$}};
\fill[color=green!50!black] (-0.5,-0.5) circle (2.9pt) node[left] {{\small$\alpha$}};
\fill[color=green!50!black]  (0,0.6) circle (2.9pt) node (meet) {};
\draw[color=green!50!black]  (-0.5,-0.5) .. controls +(0,0.5) and +(-0.5,-0.5) .. (0,0.6);
\end{tikzpicture} 
, \quad \
\beta^Q(\Phi) = 
\begin{tikzpicture}[very thick,scale=0.45,color=blue!50!black, baseline]
\nicepalecolourscheme (0,0) circle (2.25);
\shadedraw[top color=white, bottom color=white, draw=white] (0,0) circle (1.5);
\draw (0,0) circle (1.5);
\draw[->, very thick] (0.100,-1.5) -- (-0.101,-1.5) node[above] {};
\draw[->, very thick] (-0.100,1.5) -- (0.101,1.5) node[below] {}; 
\fill (202.5:1.5) circle (3.9pt) node[right] {{\small$\Phi$}};
\fill (270:1.5) circle (0pt) node[above] {{\small$Q$}};
\fill[color=green!50!black] (155:1.5) circle (3.9pt) node[right] {};
\draw[color=green!50!black] (155:1.5) .. controls +(-0.3,0.4) and +(0,-0.5) .. (135:2.5);
%
%
\end{tikzpicture} 
\qquad\text{with}\quad
\begin{tikzpicture}[very thick,scale=0.6,color=blue!50!black, baseline]

\draw (0,-1.25) node[left] (X) {};
\draw (0,1.25) node[left] (Xu) {};

\draw (0,-1.25) -- (0,1.25); 

\fill[color=green!50!black] (0,0.0) circle (2.9pt) node (meet) {};
\draw[color=green!50!black] (-0.75,1.25) .. controls +(0,-0.5) and +(-0.5,0.5) .. (0,0.0);
\end{tikzpicture} 
:=
\begin{tikzpicture}[very thick,scale=0.6,color=blue!50!black, baseline=0cm]

\draw (1.5,-1.25) -- (1.5,1.25); 

\draw (1.5,-1.25) node[left] (X) {};
\draw (1.5,1.25) node[left] (Xu) {};

\fill[color=green!50!black] (1.5,0.3) circle (2.9pt) node (meet) {};

\draw[color=green!50!black] (1,-0.25) .. controls +(0.0,0.25) and +(-0.25,-0.25) .. (1.5,0.3);
\draw[color=green!50!black] (0.5,-0.25) .. controls +(0.0,0.5) and +(0,-0.25) .. (0.5,1.25);

\draw[-dot-, color=green!50!black] (0.5,-0.25) .. controls +(0,-0.5) and +(0,-0.5) .. (1,-0.25);

\draw[color=green!50!black] (0.75,-1.2) node[Odot] (unit) {}; 
\draw[color=green!50!black] (0.75,-0.6) -- (unit);

\end{tikzpicture}
\, . 
$$
Again we note that these diagrams directly correspond to the physical picture, expressed in terms of the familiar junction fields: $\beta_Q(\alpha)$ shows how the bulk field~$\alpha$ `approaches'~$Q$ to become a boundary operator, and $\beta^Q(\Phi)$ is the bulk field that results from punching a tiny hole with boundary~$Q$ and a $\Phi$-insertion into the worldsheet. 
Distances and sizes are immaterial in topological theories. 

One finds that the restriction 
$
\beta_Q: 
\Hcc \rightarrow \End_A(Q)
$ 
is an algebra morphism with image in the centre, while the restriction 
$
\beta_Q: 
\Hrr \rightarrow \Hom_A(Q,{}_{\gamma_A}Q)
$
satisfies $\beta_Q(\alpha \cdot \alpha') = \beta_Q(\alpha) \beta_Q(\alpha')$ for $\alpha \in \Hrr$ and $\alpha' \in \Hcc$. Furthermore, $\beta^Q$ maps every element of $\End_A(Q)$ to $\Hrr$. 

The disc two-point correlator for boundary operators $\Phi_1: P \rightarrow Q$ and $\Phi_2: Q\rightarrow P$ is exactly as in the unorbifolded theory: 
$$
\big\langle \Phi_1, \Phi_2 \big\rangle_{Q,P} = 
\begin{tikzpicture}[very thick,scale=0.5,color=blue!50!black, baseline]
\nicepalecolourscheme (0,0) circle (1.5);
\draw (0,0) circle (1.5);
\draw[<-, very thick] (0.100,-1.5) -- (-0.101,-1.5) node[above] {}; 
\draw[<-, very thick] (-0.100,1.5) -- (0.101,1.5) node[below] {}; 
\fill (-22.5:1.5) circle (3.3pt) node[left] {{\small$\Phi_2$}};
\fill (22.5:1.5) circle (3.3pt) node[left] {{\small$\Phi_1$}};
\fill (270:1.5) circle (0pt) node[above] {{\small$Q$}};
\end{tikzpicture} 
\; , \quad
\big\langle -,- \big\rangle_{Q,Q} \equiv \big\langle -,- \big\rangle_{Q} \, .
$$
This pairing can be shown to be nondegenerate when restricted to 
$
\Hom_A(P,Q) \times \Hom_A(Q,{}_{\gamma_A}P) 
$. 
Put differently, the Serre functor on the category of boundary conditions in generalised orbifolds is given by ${}_{\gamma_A}(-)$, i.\,e.~twisting with the Nakayama automorphism. 

The compatibility between $\beta_Q, \beta^Q$ and the two-point correlators is 
\begin{align*}
\Big\langle \alpha, \; \beta^Q(\Phi) \Big\rangle_{(a,A)} = \Big \langle \beta_Q(\alpha),\; \Phi \Big\rangle_Q \, ,  \quad 
\Big\langle \beta^Q(\Phi),\; \alpha \Big\rangle_{(a,A)} = \Big \langle \Phi,\; \beta_Q(\gamma_A \circ \alpha) \Big\rangle_Q \, . 
\end{align*}
This is the familiar adjunction between $\beta_Q$ and $\beta^Q$, but only up to another `twist' by~$\gamma_A$. 
Finally, there is a version of open/closed duality known as the \textsl{Cardy condition}, which asserts that the overlap of two boundary states $\beta^Q(\Phi)$, $\beta^{P}(\Psi)$ computed in the closed sector is equal to an open sector `partition function' or trace. The precise identity is 
$$
\Big\langle \beta^Q(\Phi) \cdot \beta^{P}(\Psi) \Big\rangle_{(a,A)}
=
\tr({}_\Psi m_\Phi)
$$
where $\Phi \in \End_A(Q)$, $\Psi \in \End_A(P)$, and the trace of ${}_\Psi m_\Phi: \xi \mapsto \Psi\xi\Phi$ is taken over the $\C$-vector space $\Hom_A(Q,P)$. In particular, the proof of the Cardy condition holds for the special case of ordinary $G$-orbifolds, i.\,e.~for $A = A_G$. 

\medskip

We now turn to some basic properties of the defect sector of generalised orbifolds. Given two theories $(a,A)$ and $(b,B)$ and a defect~$X$ between them, a basic question is how~$X$ acts on bulk fields. The general answer is always `by wrapping the defect around the field insertion' which here means that the (right) action $\mathcal D_{\text{r}}(X)$ of~$X$ on a twisted bulk field $\alpha \in \Hom(I_a,A)$ produces the bulk field 
$$
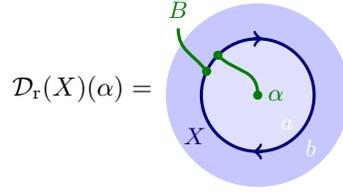
\mathcal D_{\text{r}}(X)(\alpha) = 
\begin{tikzpicture}[very thick,scale=0.6,color=blue!50!black, baseline]
\nicecolourscheme (0,0) circle (2.0);
\fill (1.175,-1.175) circle (0pt) node[white] {{\small$b$}};
\nicepalecolourscheme (0,0) circle (1.25);
\fill (0.65,-0.65) circle (0pt) node[white] {{\small$a$}};
\draw (0,0) circle (1.25);
\fill (-135:1.3) circle (0pt) node[left] {{\small$X$}};
\draw[->, very thick] (0.100,-1.25) -- (-0.101,-1.25) node[above] {}; 
\draw[->, very thick] (-0.100,1.25) -- (0.101,1.25) node[below] {}; 
\fill[color=green!50!black] (135:0) circle (2.9pt) node[right] {{\small$\alpha$}};
\draw[color=green!50!black] (0,0) .. controls +(0,0.6) and +(0.4,-0.4) .. (135:1.25);
\fill[color=green!50!black] (135:1.25) circle (2.9pt) node[right] {};
\fill[color=green!50!black] (155:1.25) circle (2.9pt) node[right] {};
\draw[color=green!50!black] (155:1.25) .. controls +(-0.3,0.4) and +(0,-0.5) .. (140:2.3);
\fill[color=green!50!black] (140:2.3) circle (0pt) node[above] {{\small$B$}};
\end{tikzpicture} 
$$
in the theory $(b,B)$. Similarly one has a left action $\mathcal D_{\text{l}}(X): \Hom(I_b,B) \rightarrow \Hom(I_a,A)$ by wrapping~$X$ counterclockwise. These defect actions induce operators 
$$
\mathcal D_{\text{l}}(X): \mathcal H_{\text{RR}}^B \lra \mathcal H_{\text{RR}}^A
\, , \quad
\mathcal D_{\text{r}}(X): \mathcal H_{\text{RR}}^A \lra \mathcal H_{\text{RR}}^B
$$
between the respective spaces of orbifold projected RR ground states $\mathcal H_{\text{RR}}^A, \mathcal H_{\text{RR}}^B$. The latter are defined as in~\eqref{eq:HRRspace} with $A_G$ replaced by $A,B$, respectively. 

As expected the invisible defect acts as the identity, i.\,e.~$\mathcal D_{\text{l}}(A) = \mathcal D_{\text{r}}(A) = 1$ on $\Hrr$, and orientation reversal is the same as wrapping the defect in the opposite direction, $\mathcal D_{\text{l}}(X) = \mathcal D_{\text{r}}(X^\dagger)$. Furthermore, defect action is compatible with fusion: if $Y: (b,B) \rightarrow (c,C)$ is another defect we have 
$$
\mathcal D_{\text{l}}(X) \circ \mathcal D_{\text{l}}(Y) = \mathcal D_{\text{l}}(Y\otimes_B X)
\, , \quad
\mathcal D_{\text{r}}(Y) \circ \mathcal D_{\text{r}}(X) = \mathcal D_{\text{r}}(Y\otimes_B X)
\, . 
$$
Here we encounter the fact that defect fusion in generalised orbifolds is described by the tensor product $\otimes_B$ over the algebra~$B$ encoding the intermediate theory $(b,B)$ that is `squeezed out' in the fusion process. This tensor product $Y\otimes_B X$ is defined by a universal property; it is a subsystem of the fusion $Y\otimes X$ in the unorbifolded theory and can typically be computed as the image of an explicit projector, see e.\,g.~\cite[Sect.\,3]{genorb}. 

Finally the left and right defect actions are adjoint with respect to the bulk pairings, again up to a twist of the Nakayama automorphism: 
\begin{align*}
\Big\langle \beta, \mathcal D_{\text{r}}(X) (\alpha) \Big\rangle_{(b,B)}
& = 
\Big\langle \mathcal D_{\text{l}}(X) (\beta) , \alpha \Big\rangle_{(a,A)} 
\, ,  \\
\Big\langle \alpha, \mathcal D_{\text{l}}(X) (\beta) \Big\rangle_{(a,A)}
& = 
\Big\langle \mathcal D_{\text{r}}(X) (\gamma^{-1}_A \circ \alpha) , \gamma^{-1}_B \circ \beta \Big\rangle_{(b,B)} 
\end{align*}
for all $\alpha \in \Hrr$ and $\beta \in \mathcal{H}_{\text{RR}}^B$. This captures the intuitive idea that a defect loop around one field insertion on a sphere can be pulled around to wrap another field insertion in the opposite direction. 

\begin{acknowledgements}
\textup{We thank Dan Murfet, Ingo Runkel and Yuji Tachikawa for helpful discussions.} 
\end{acknowledgements}

\end{document}